\documentclass[fleqn,10pt]{wlscirep}

\usepackage{epstopdf}
\usepackage[english]{babel}
\usepackage{graphicx}

\newcommand{\Cref}[1]{Chapter~\ref{#1}}

\newcommand{\Eref}[1]{Equation~\eqref{#1}}

\newcommand{\Fref}[1]{Figure~\ref{#1}}
\newcommand{\Fsref}[1]{Figures~\ref{#1}}

\newcommand{\Rref}[1]{Reference~\citenum{#1}}

\title{Point-dipole approximation
 for small systems of strongly coupled
 radiating nanorods}

\author[1*]{Derek W. Watson}
\author[1]{Stewart D. Jenkins}
\author[2]{Vassili A. Fedotov}
\author[1,3]{Janne Ruostekoski}
\affil[1]{Mathematical Sciences and
 Centre for Photonic Metamaterials,
 University of Southampton,
 Southampton SO17 1BJ,
 United Kingdom}
\affil[2]{Optoelectronics Research Centre and
 Centre for Photonic  Metamaterials,
 University of Southampton, Southampton SO17 1BJ,
 United Kingdom}
\affil[3]{Department of Physics, Lancaster
 University, Lancaster, LA1 4YB,
 United Kingdom}

\affil{*derek.w.watson@gmail.com}

\begin{abstract}
 Systems of closely-spaced resonators
 can be strongly coupled by interactions
 mediated by scattered electromagnetic fields.
 In large systems the resulting
 response has been shown
 to be more sensitive to these 
 collective interactions than to the detailed
 structure of individual resonators.  
 Attempts to describe such systems
 have resulted in point-dipole approximations
 to resonators that are computationally 
 efficient for large resonator ensembles. Here
 we provide a detailed study for the validity 
 of point dipole approximations in small systems
 of strongly coupled plasmonic nanorods, 
 including the cases of both super-radiant and subradiant
 excitations, where the 
 characteristics of the excitation
 depends on the spatial separation between the nanorods.
 We show that over an appreciable range of rod lengths 
 centered on
 $210~\text{nm}$, 
 when the relative separation $kl$ in terms of the 
 resonance wave number of light $k$
 satisfies $kl \gtrsim \pi/2$, the point electric
 dipole model becomes accurate. 
 However, when the resonators are closer, the
 finite-size and geometry of the resonators
 modifies the excitation modes, in particular the 
 cooperative mode line shifts of the point dipole 
 approximation begin to rapidly
 diverge at small separations. 
 We also construct simplified effective models by describing
 a pair of nanorods as a single effective
 metamolecule.
\end{abstract}
\begin{document}

\flushbottom
\maketitle

\thispagestyle{empty}

 \section*{Introduction}
 \label{sec:intro}

 Plasmonic nanorods are the most basic form 
 of optical resonators. 
 The scattering of light from any resonator has the ability
 to produce strong interactions that can result from the
 wave repeatedly scattering off the same resonator.
 The EM coupling
 between different resonators results in different
 eigenmodes of response, the strong interaction limits of 
 which can be most easily achieved
 in microwave systems with low 
 ohmic losses,
 but can also manifest in plasmonic 
 systems~\cite{PhysRevLett.119.053901,
 PhysRevB.98.245136}. 
 The eigenmodes
 can also destructively interfere and manifest
 as Fano resonances in the transmitted field~\cite{Liu,
 doi:10.1021/nn401175j,
 Fan1135}, whose narrow resonances
 potentially make them
 useful in applications such as
 plasmonic rulers~\cite{Liu1407}
 and biosensors~\cite{doi:10.1021/nl102991v}.
 Designing material structures
 to support  Fano  resonances  is difficult;
 not least due to the complex
 interactions of different modes, but variations in
 the resonators
 can affect the line shifts and widths of
 the interacting modes also~\cite{doi:10.1021/nl1032588}.

 Apart from 
 applications in plasmonics and nanophotonics,
 optical resonators (and nanorods in particular) that 
 are much smaller than the wavelength of the driving 
 light are now 
 commonly used as the building blocks of 
 metamaterials - artificial material composites that are 
 designed to interact with light in ways no conventional 
 materials can. 
 Functionalities of
 metamaterials include
 perfect absorption~\cite{PhysRevLett.100.207402}
 and optical magnetism~\cite{NatPonShalaev}; with
 potential  applications ranging from
 cloaking~\cite{1236073,
 Schurig10112006,
 Pendry23062006} to perfect
 lenses~\cite{PhysRevLett.85.3966,
 Jacob:06,
 PhysRevLett.92.117403}.
 An important part of understanding how 
 metamaterials realize their functions is knowledge of the
 electromagnetic (EM) fields scattered by the 
 constituent resonators.

 Metamaterials that exhibit strong collective
 interactions are becoming increasingly
 popular with experimentalist~\cite{PhysRevLett.104.223901,
 PhysRevB.80.041102,
 LemoultPRL10,
 ValentineNatComm2014,
 PhysRevLett.119.053901,
 Anlageprx,
 PhysRevE.95.050201,
 doi:10.1063/1.3138868}.
 However, modeling the EM interactions
 in large resonator systems is generally
 challenging. The interactions among the 
 resonators can be
 simplified, e.g., by treating  the array as an
 infinite lattice~\cite{PhysRevLett.84.4184,
 798002}
or the resonators as point multipole
 sources~\cite{PhysRevB.82.075102,
 PhysRevLett.119.053901,
 PhysRevB.86.085116,
 PhysRevA.95.033822}.
 Point multipole descriptions in particular have
 been
 successful in modeling the
 cooperative response in planar arrays, e.g.,
 developing electron-beam-driven light 
 sources~\cite{PhysRevLett.109.217401} and 
 transmission properties~\cite{JenkinsLineWidthNJP,
 PhysRevLett.111.147401}.
 Point dipole descriptions can also be
 extended to more complex metamolecules,
 such as those exhibiting
 toroidal dipoles~\cite{PhysRevB.93.125420}.

 Here, we analyze the accuracy of the dipole approximation
 in more detail in small systems 
 of plasmonic nanorods and show, qualitatively, at what
 separations the finite-size of a nanorod, and
 its near fields, must be accounted for.
 Our theoretical model does not require solving the
 full Maxwell's equations for a
resonator ensemble; which is computationally demanding for
 more than a few resonators~\cite{5565504}.
 Rather, we utilize the formalism developed
 in \Rref{PhysRevB.86.085116}
 to produce a system of coupled equations
 for the dynamics of the EM
 interactions of the scattered and incident EM fields.
 The method relies on capturing the fundamental
 physics of each resonator, e.g., its decay rate and resonance
 frequency, relevant for the radiative coupling
 between resonators. Our model is readily applied
 to more complex systems such as those whose resonators 
 are distributed over two planes (one above another) 
 with non-uniform 
 orientations, e.g., a
 toroidal metamolecule~\cite{PhysRevB.93.125420}.
 
 Finally, in our paper we also provide an alternative approach
 for treating each element of a nanorod as
 a separate meta-atom
 when we model closely spaced
 nanorods as a single effective metamolecule.
 This can notably reduce the number
 of degrees of freedom in the system.

 \section*{Method}
 \label{sec:method}

 In analyzing the EM interactions between
 plasmonic nanorods and an incident EM field, we utilize the
 general theory derived in detail in 
 \Rref{PhysRevB.86.085116},  
 specifically, for the point electric dipole approximation.
 We regard nanorods as cylinder-shaped 
 resonators and 
 study their longitudinal polarization excitation;
 where the charge
 and current oscillations are assumed
 to be linear along the axis of the nanorod.
 Here, we provide a brief overview
 of our model, a more detailed description is provided in the
 Supplementary Material.

 An incident electric displacement field
 ${\bf D}_{\text{in}}({\bf r},t)
 =
 D_{\text{in}}{\bf \hat e}_{\text{in}}\,
 \exp(i{\bf k}_{\text{in}}\cdotp {\bf r} - i\Omega_0t)$
 and magnetic induction
 ${\bf B}_{\text{in}}({\bf r},t) = \sqrt{\mu_0/\epsilon_0}\,
 {\bf \hat k}_{\text{in}}\times{\bf E}_{\text{in}}({\bf r},t)$,
 with
 frequency $\Omega_0$, polarization vector
 ${\bf \hat e}_\text{in}$  and
 propagation vector ${\bf k}_\text{in}$
 drive each resonator $j$'s
 internal charge and current sources.
 These source oscillations 
 behave in a manner comparable to 
 an LC circuit with resonance frequency 
 $\omega_j$~\cite{PhysRevB.86.085116},
 \begin{equation}
 \omega_j = \frac{1}{\sqrt{L_jC_j}}
 \,\textrm{.}
 \label{eq:omega}
 \end{equation}
 Here, $L_j$ and $C_j$ are respectively,
 an effective self-inductance 
 and self-capacitance.
 A dynamic variable
 with units of charge
 $Q_j(t)$ and its time derivative, the 
 current $I_j(t) = \dot {Q_j}(t)$, describes the 
 state of current oscillations within each resonator $j$.
 In order to analyze the coupled equations for the
 EM fields, we introduce
 the slowly varying normal mode oscillator
 amplitudes $b_j(t)$~\cite{PhysRevB.86.085116},
 with generalized coordinate the charge $Q_j(t)$ 
 and conjugate momentum $\phi_j(t)$, where
 \begin{equation}
 b_j(t) = \frac{1}{\sqrt{2\omega_j}}
 \left[
 \frac{Q_j(t)}{\sqrt{C_j}} + i \frac{\phi_j(t)}{\sqrt{L_j}}
 \right]
 \,\textrm{.}
 \label{eq:b}
 \end{equation}
 In the rotating wave approximation
 the conjugate momentum  and current are
 linearly-proportional~\cite{PhysRevB.86.085116}.
 We use \Eref{eq:b} to describe a general resonator
 with sources of both polarization ${\bf P}_j({\bf r},t)$ and
 magnetization ${\bf M}_j({\bf r},t)$.
 The resonator's scattered EM fields  result from 
 the oscillations of $Q_j(t)$ and $I_j(t)$; which are 
 proportional to ${\bf P}_j({\bf r},t)$  and
 ${\bf M}_j({\bf r},t)$,
 respectively~\cite{PhysRevB.86.085116}.

 The
 collective interactions of $N$ resonators
 with each other and an external field
 is described by the linear system 
 of equations\cite{PhysRevB.86.085116}
 \begin{equation}
 {\bf \dot b} = \mathcal{C}{\bf b} + {\bf F}_{\text{in}}
 \,\textrm{.}
 \label{eq:equmot}
 \end{equation}
 Here, ${\bf \dot{b}}$ is the rate  
 of change of ${\bf b}$; a vector of $N$ normal
 oscillator variables, and ${\bf F}_{\text{in}}$
 is a vector describing the interaction of
 resonator $j$ with the incident EM field.
 The $N\times N$ interaction 
 matrix $\mathcal{C}$ requires solving the
 scattered  EM fields
 for the polarization and magnetization sources.
 The EM interactions between different resonators
 $i \ne j$
 are described by the 
 off-diagonal elements 
 of $\mathcal{C}$. The 
 strength of $\mathcal{C}_{i\ne j}$
 depends on the separation and
 orientation of the  resonators,
  and naturally 
 accounts for the reduced coupling 
 between elements at the edges compared 
 to the center of a lattice,
 see Supplementary Material.
 The diagonal elements describe the
 EM interactions of a resonator with 
 itself, resulting in the resonator's
 decay rate $\Gamma_j$ and resonance 
 frequency;
 \begin{equation}
 \begin{bmatrix}
 \mathcal{C}
 \end{bmatrix}_{jj}
 =
 - \frac{\Gamma_j}{2}  -i(\omega_j - \Omega_0)
 \,\text{.}
 \label{eq:C_jj}
 \end{equation}
 
 In our model we consider magnetization
 due to induced macroscopic currents.
 In a straight rod even though the
 induced current is non-zero, it is
 linear and therefore the corresponding
 magnetization is negligible;
 ${\bf M}_j({\bf r},t)\simeq 0$.
 Thus, the scattered EM fields result from a nanorod's
 polarization sources
 ${\bf P}_j({\bf r},t)$ alone. This results in 
 an effective accumulation of charge
 on the nanorod's ends.
 Analogous simulation methods can also be 
 used to study collective responses of arrays 
 of other resonant emitters,
 such as atoms~\cite{PhysRevLett.117.243601}.

  \subsection*{Finite-size resonator model}
  
  \begin{figure}[h!]
 \centering
 \includegraphics[]{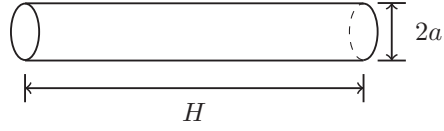}
 \caption[Geometry of a single nanorod.]
 {\label{fig:1}Schematic of a nanorod
 with radius $a$ and length $H$.
 }
 \end{figure}

 For our finite-size model we consider a nanorod
 with a radius $a$ and height $H_j$, see \Fref{fig:1}.
 The polarization
 density is a uniform
 distribution of atomic point electric dipoles
 (with orientation vectors ${\bf \hat d}_j$)
 throughout the volume
 of the nanorod. For a single nanorod centered at the origin
 and aligned along the $z$ axis,
 i.e., ${\bf \hat d}_j = {\bf \hat z}$,
 the spatial profile distribution of the
 polarization density is
 \begin{equation}
 {\bf P}_j({\bf r},t)
 =
 \frac{Q_j(t)}{\pi a^2}{\bf \hat z}
 \Theta(a - \rho)
 \Theta(H_j/2-z)
 \Theta(H_j/2+z)
 \,\text{,}
 \label{eq:p_finite}
 \end{equation}
 where $\Theta$ is the Heaviside function and $\rho < a$.
 An individual nanorod experiences radiative
 decay $\Gamma_{\text{E1},j}$, resulting
 from the interaction of the scattered field
 with the rod itself. The total decay rate of the rod
 $\Gamma_j$ also includes a
 phenomenological ohmic loss
 rate $\Gamma_{\text{O},j}$,
 where
 \begin{equation}
 \Gamma_j =
 \Gamma_{\text{E1},j}
 +
 \Gamma_{\text{O},j}
 \,\text{,}
 \quad
  \Gamma_{\text{E1},j}
 =\frac{C_jH_j^2\omega_j^4}{6\pi\epsilon_0c^3}
 \,\text{.}
 \label{eq:GammaE1}
 \end{equation}

 \subsection*{Point electric dipole approximation}

 In a point multipole approximation
 where the length is much less than the wavelength
 of the incident light ($H_j\ll \lambda_0$), the nanorod
 may be approximated as a point
 electric dipole with polarization
 density
 \begin{equation}
 {\bf P}_j({\bf r},t) =
 Q_j(t)H_j{\bf \hat d}_j\delta({\bf r} - {\bf r}_j)
 \,\text{.}
 \label{eq:PE1}
 \end{equation}
 The orientation of the electric dipole
 is described by the unit vector ${\bf \hat d}_j$,
 and $H_j$ has dimensions of length.
 The rate at which the electric
 dipole radiates as a result of its self
 interactions is $\Gamma_{\text{E1},j}$.  
 The radiative decay, ohmic losses and total decay rate 
 are given in \Eref{eq:GammaE1}.

 \subsection*{A finite-size effective metamolecule}
 
 Thus far, we have only considered the
 interactions between individual 
 nanorods and point electric dipole resonators. 
 When there are a large number of
 resonators, one may look to  
 optimize the model. Because resonators are often arranged
 in a lattice framework, it is possible to
 consider closely spaced parallel 
 pairs of nanorods
 as a single effective metamolecule. 
 Here, we 
 extend our finite-size nanorod model
 to include this effective metamolecule with the same
 properties as its constituent nanorods,
 i.e., electric dipole properties.

 We consider two parallel nanorods with location 
 vectors
 ${\bf r}_{\pm,j} = [x_j, y_j\pm l,z_j]$, and
 polarization densities
 ${\bf P}_{\pm,j}({\bf r},t)$, respectively.
 When $l$ is small, we may 
 approximate the pair as a single metamolecule
 with location vector ${\bf r}_j = [x_j,y_j,z_j]$.
 The metamolecule may be symmetrically excited,
 i.e., ${\bf P}_{+,j}({\bf r},t)
 = {\bf P}_{-,j}({\bf r},t)$, or it may
 be antisymmetrically excited,
 i.e., ${\bf P}_{+,j}({\bf r},t)
 = -{\bf P}_{-,j}({\bf r},t)$, where
 ${\bf P}_{+,j}({\bf r},t)$ is defined in \Eref{eq:p_finite}.

 The  resonance frequency 
 and radiative decay rate of our 
 effective metamolecule depend on the interactions
 of the individual nanorods.
 We calculate these properties later;
 by analyzing in detail
 a pair of parallel nanorods and point electric dipoles.
 In principle, if the nanorods are approximated
 as point electric dipoles, one may treat  a
 closely spaced parallel pair as a single emitter. If both
 the electric dipoles are symmetrically 
 excited, there is an effective single point electric
 dipole~\cite{PhysRevB.96.035403}. However, 
 if both the dipoles are antisymmetrically excited, 
 there is an effective emitter with both an electric quadrupole and a 
 magnetic dipole moment~\cite{PhysRevB.96.035403}. In this work,
 however, we only consider the finite-size effective resonators.

 \section*{Results}
 \label{sec:results}

 In this section, we analyze the EM
 interactions of nanorods,
 both as point electric dipole emitters
 and accounting for their
 finite-size and geometry.
 We formulate the model for 
 the nanorods by assuming all the
 nanorods are made of gold and have equal length, 
 i.e., $H_j = H_0$ for all $j$. 
 As the
 nanorods are identical, they experience
 identical ohmic losses, radiative,
 and total decay rates, i.e.,
 $\Gamma_{\text{O},j} = \Gamma_{\text{O}}$,
 $\Gamma_{\text{E1},j}=\Gamma_{\text{E1}}$,
  and
 $\Gamma_j=\Gamma$. 
 We choose
 $H_0 = 1.5\lambda_{\text{p}}\simeq 210~\text{nm}$ and
 radius
 $a = \lambda_p/5\simeq 28~\text{nm}$, where
 $\lambda_{\text{p}} \simeq 140~\text{nm}$ is the
 plasma wavelength of
 gold~\cite{doi:10.1021/j100287a028Schatz}.
 This yields
 $H_0 \simeq 0.24\lambda_0$ and
 $a \simeq 0.032\lambda_0$, where
 $\lambda_0 = 2\pi c/\omega_0
 \simeq 860~\text{nm}$ is the (longitudinal)
 resonance wavelength of the nanorod.
 Each individual
 nanorod
 has a total decay rate $\Gamma$,
 with resulting radiative emission rate
 $\Gamma_{\text{E1}} \simeq 0.83\Gamma$
 and ohmic loss rate
 $\Gamma_{\text{O}} \simeq 0.17\Gamma$. 
 We  calculate these parameters in the
 Supplementary Material, where we employ 
 formulas developed in 
 \Rref{kuwate} for the resonant scattering of 
 light from plasmonic nanoparticles, and
 ohmic losses are accounted for in the Drude model.
 To simplify the comparison
 between our point dipole approximation 
 and finite-size nanorod model, we use
 the same resonance frequency, radiative decay, ohmic losses 
 and total decay rate in both models.

 We analyze the characteristic response 
 of the system in the absense of an incident 
 EM field, studying the characteristic 
 collective modes represented
 by the eigenvectors ${\bf v}_n$ of 
 the interaction matrix $\mathcal{C}$. 
 The corresponding complex eigenvalues $\xi_n$,
 describe the collective mode's 
 characteristic linewidth (real part) 
 and resonance frequency shift 
 (imaginary part);
 $\xi_n =
 - \gamma_n/2
  -i(\Omega_n - \omega_0)$.
 The number of modes is determined 
 by the number of resonators, 
 and the radiation 
 may be suppressed $\gamma_n < \Gamma$
 (subradiant), or enhanced 
 $\gamma_n > \Gamma$ (superradiant).

 \subsection*{Two parallel nanorods}
 \label{sec:twoNanorods}

 As our first example, we consider two
 parallel nanorods (and two point electric dipoles)
 with location vectors
 ${\bf r}_1 = -{\bf r}_2 = [0,l/2,0]$.
 The coupling matrix $\mathcal{C}$
 has two eigenmodes of current oscillation. In
 the first mode, the current oscillations are 
 in-phase (symmetric) with 
 ${\bf \hat d}_1 = {\bf \hat d}_2$. 
 In the second mode,  the current 
 oscillations are out-of-phase 
 (antisymmetric) with ${\bf \hat d}_1 = -{\bf \hat d}_2$.
 The former we denote by a subscript `s'; and
 latter by a subscript `a'. 

\begin{figure}[h!]
 \centering
 \includegraphics[]{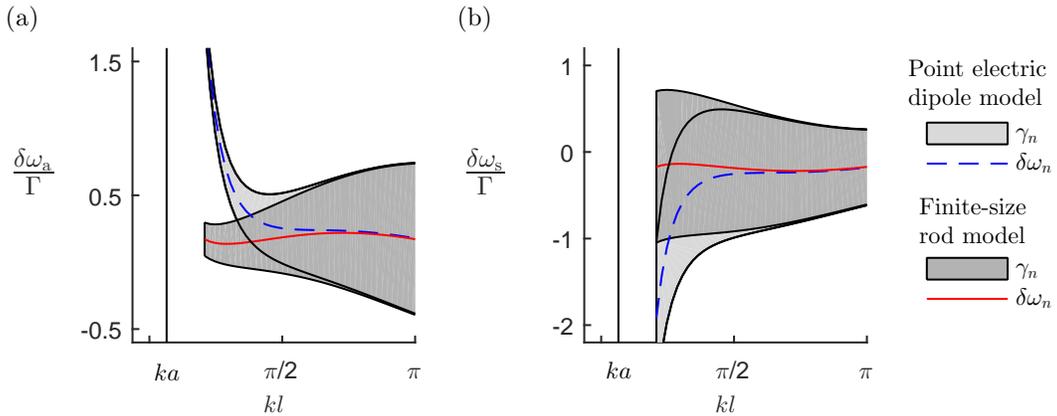}
 \caption
 {\label{fig:2}
 The radiative resonance 
 line shifts
 $\delta\omega_n = -(\Omega_n - \omega_0)$, 
 and linewidths $\gamma_n$,
 for the collective out-of-phase~(a)
 and in-phase~(b) eigenmodes, for two 
 parallel nanorods and  two parallel point electric dipoles as a
 function of their separation
 $l$.
 The finite-size rods have
 lengths $H\simeq0.24\lambda_0$ and
 radii $a\simeq0.032\lambda_0$. The radiative 
 decay rate of
 each nanorod is
 $\Gamma_{\text{E1}}\simeq0.83\Gamma$,
 the ohmic losses are
 $\Gamma_{\text{O}}\simeq0.17\Gamma$.
 }
 \end{figure}

 In \Fref{fig:2}, we show the
 collective eigenmode's 
 radiative resonance  line shifts and linewidths
 for two
 finite-size parallel nanorods and
 two parallel point electric dipoles. 
 Throughout the range of $kl$, the decay rates
 $\gamma_n$ of the point dipole
 model closely agree with
 the corresponding decay rates of
 the finite-size nanorod model.
 When $kl \approx \pi$, the symmetric mode is 
 subradiant with $\gamma_{\text{s}}\approx 0.9\Gamma$, 
 and the antisymmetric mode is superradiant with
 $\gamma_{\text{a}} \approx 1.1\Gamma$;
 for both the nanorods and point electric dipoles. As the 
 separation reduces, the symmetric
 mode becomes superradiant
 and the antisymmetric mode subradiant.
 When $kl\approx 2\pi/3$, the decay rates
 of the point multipole approximation are:
 $\gamma_{\text{a}}
 \approx 0.7\Gamma$
 (subradiant);
 and $\gamma_{\text{s}}
 \approx 1.3\Gamma$
 (superradiant); the finite-size model shows decay rates:
 $\gamma_{\text{a}}
 \approx 0.8\Gamma$;
 and
 $\gamma_{\text{s}}
 \approx 1.2\Gamma$.
 As the separation becomes small $kl\approx\pi/6$, the
 antisymmetric linewidths approach the
 ohmic loss rate,
 $\gamma_{\text{a}}
 \approx 0.2\Gamma$
 and the symmetric mode
 linewidths become more superradiant
 $\gamma_{\text{s}}
 \approx 1.8\Gamma$.

 The lineshifts $\delta\omega_n$, however,
 only agree when $kl\gtrsim\pi/2$.
 As the separation becomes small $kl\lesssim\pi/2$, the
 line shift of the point electric dipole model begins to
 separate from the corresponding line shift
 of the finite-size resonator model.
 The line shift of the finite-size resonator model here
 is   $\Omega_{\text{a}}^{\text{(1)}} - \omega_0 =
 -(\Omega_{\text{s}}^{\text{(1)}} - \omega_0)
 \simeq 2.5\Gamma$.
 For $kl\lesssim\pi/2$,  the antisymmetric mode line shift 
 of the point electric dipole
 model is blue shifted from $\omega_0$,
 and the symmetric mode  red shifted.

 In \Fref{fig:2}, the nanorods
 have length $H_0 \simeq 0.24\lambda_0$ and 
 radius $a=\lambda_\text{p}/5$.
 In the Supplementary Material, we show the 
 line shifts and linewidths when driven 
 on resonance for  longer and shorter rods,
 still with radius $a = \lambda_\text{p}/5$.
 The longer rods have length
 $H_{l}=2H_0 \simeq 0.27\lambda_{l}$, where 
 $\lambda_{l}\simeq 1540~\text{nm}$ is the resonance 
 wavelength of the longer rod, and shorter rods
 $H_{t} = H_0/2\simeq 0.18\lambda_{t}$, where
 $\lambda_{t}\simeq 570~\text{nm}$.
 For both longer and shorter rod systems, the 
 point dipole approximation becomes valid when
 the separation is $kl\gtrsim\pi/2$, where $k$
 is the resonance wavenumber 
 of the light. 
 When we make small changes to the nanorod
 radius $\lambda_\text{p}/6 < a < \lambda_\text{p}/4$;
 we also find the point
 dipole approximation is valid for separations
  $kl\gtrsim \pi/2$.

 \subsection*{Two interacting pairs of nanorods}
 \label{sec:twoPairs}

 In this section, we analyze
 two parallel pairs of horizontal nanorods. Firstly,
 we treat the interacting pairs as four
 discrete finite-size nanorods that
 have a non-vanishing polarization density.
 Secondly, we optimize the model by treating the two pairs
 as two effective metamolecules. In each case,
 we compare the model to four discrete
 point electric dipoles.

 \begin{figure}[h!]
 \centering
 \includegraphics[]
 {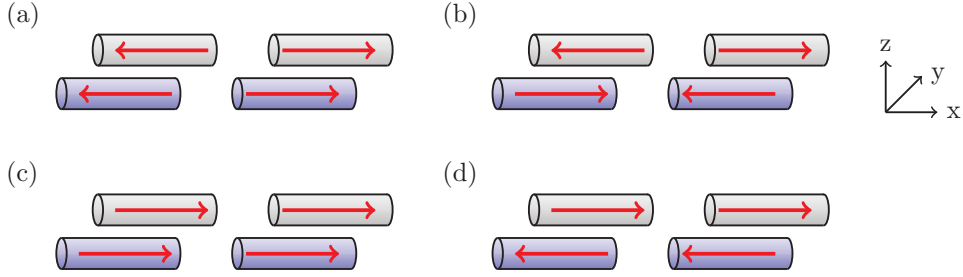}
 \caption
 {\label{fig:3}
 The eigenmodes of two horizontal
 pairs of nanorods (electric dipoles) 
 shown schematically.
 The red arrows indicate the phase of
 current oscillations, the
 shading indicates nanorods in a shared plane.
 The nanorods are located at ${\bf r}_{\pm, 1} = 
 [\lambda_0/2,\pm l/2, 0]$ and ${\bf r}_{\pm, 2} = 
 [-\lambda_0/2,\pm l/2, 0]$, where $\lambda_0$ is the resonance
 wavelength of a single nanorod.
 The modes are designated as follows:
 (a)~E1a; (b)~E2a; (c)~E1s;
 and (d)~E2s.
 }
 \end{figure}

 \subsubsection*{Four discrete nanorods}
 \label{sec:resultsFourNanorods}

 In general, we position the
 $j$th pair of nanorods at ${\bf r}_{\pm, j}
 = [x_j, y_j\pm l/2, z_j]$. In our example, we vary 
 the parameter $l$ and set
 $y_j = z_j = 0$ for each $j$ and $kx_1=-kx_2= \pi$.
 When the
 interactions between individual
 nanorods are considered, there
 are four  collective modes, see \Fref{fig:3}.
 When each parallel pair of
 nanorods are symmetrically excited,
 they can be approximated as
 out-of-phase (E1a) and in-phase (E1s)
 effective electric dipoles, see
 \Fsref{fig:3}(a)
 and~\ref{fig:3}(c), respectively.
 When the nanorods in each pair are antisymmetrically
 excited, they can be likened to two resonators with both
 electric quadrupole and magnetic dipole moments, where
 each pair are out-of-phase (E2a) or in-phase (E2s),
 see \Fsref{fig:3}(b)
 and~\ref{fig:3}(d), respectively.

 \begin{figure}[h!]
 \centering
 \includegraphics[]
 {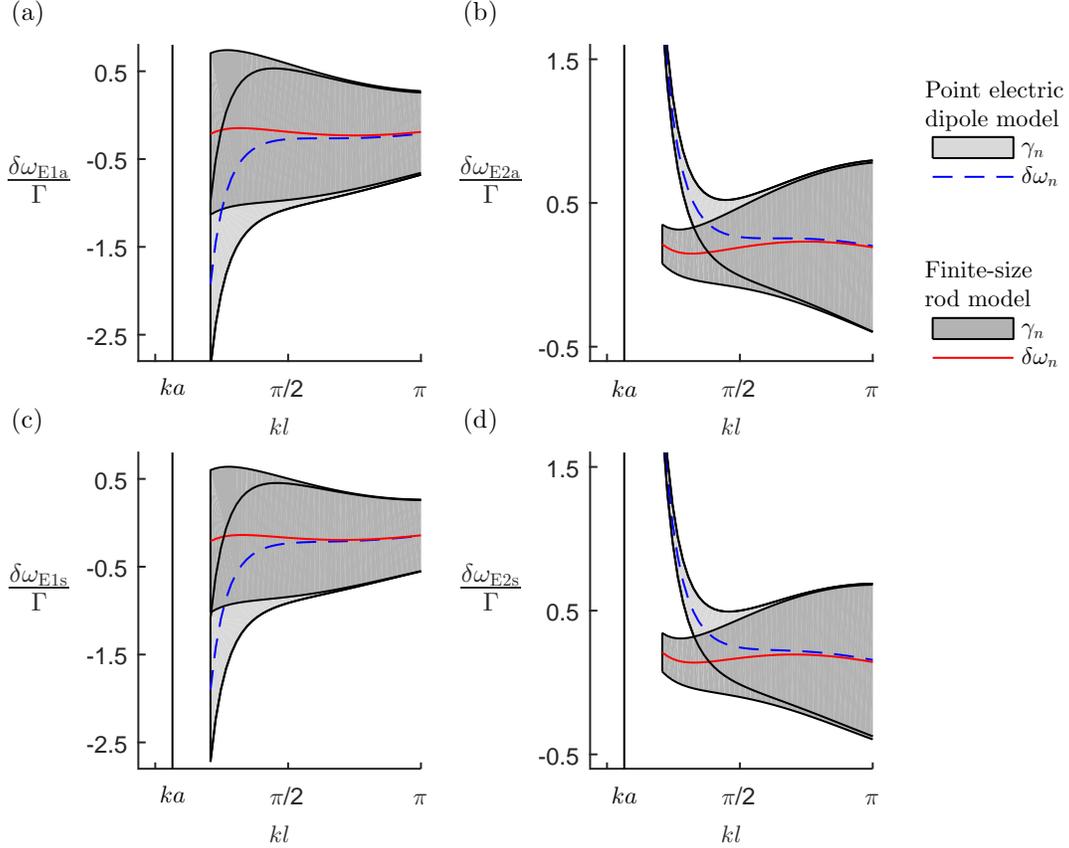}
 \caption
 {\label{fig:4}
 The radiative resonance line shifts 
 $\delta\omega = -(\Omega_n - \omega_0)$ 
 and linewidths $\gamma_n$, of two pairs of horizontal 
 nanorods and two pairs of point electric dipoles 
 located at ${\bf r}_{\pm,1} = [\lambda_0/2,\pm l/2,0]$ 
 and ${\bf r}_{\pm,2} = [-\lambda_0/2,\pm l/2,0]$, as a
 function of the parameter $l$, see \Fref{fig:3}.
 We show the collective modes: (a)~E1a; (b)~E2a; (c)~E1s;
 and (d)~E2s.
 For the nanorod parameters and plot descriptions
 see \Fref{fig:2} caption.
 }
 \end{figure}

 In \Fref{fig:4},
 we show the collective mode resonance 
 line shifts and linewidths
 of four point
 electric dipole resonators
 and those of four interacting finite-size
 nanorods. Again, we find the collective mode decay
 rates of the different models
 qualitatively agree throughout the
 range of $kl$; while
 the line shifts only agree when $kl\gtrsim\pi/2$.
 The E1a and E1s modes behave very similarly,
 as do the E2a and E2s modes.
 The deviation of the point electric dipole approximation's 
 line shift from the
 nanorod model's, begins to occur
 when the separation between parallel
 pairs of rods is $kl\simeq \pi/2$.
 Here, both the E1a and E1s modes are superradiant with
 $\gamma_{\text{E1a}}
 \simeq 1.6\Gamma$ and
 $\gamma_{\text{E1a}}
 \simeq 1.4\Gamma$.
 Conversely, here, the E2a and E2s modes are subradiant
 with
 $\gamma_{\text{E2a}}
 \simeq
 \gamma_{\text{E2s}}
 \simeq 0.6\Gamma$.

 At $kl\simeq 9\pi/10$, the E1s
 mode becomes noticeably subradiant with
 $\gamma_{\text{E1s}}\simeq 0.9\Gamma$. Here also,
 the E2a mode becomes noticeably superradiant with
 $\gamma_{\text{E2a}}\simeq 1.1\Gamma$. The E1a mode
 remains superradiant while
 the E2s mode remains subradiant
 throughout the range.

 \subsubsection*{Two effective metamolecules}
 \label{sec:resultsTwoEffectiveResonators}

 When we approximate each pair
 of nanorods as an effective single resonator; if 
 the nanorods' current oscillations  are in-phase there is 
 an effective electric dipole resonator,
 if the current oscillations
 are out-of-phase there
 is an effective resonator with both electric 
 quadrupole and  magnetic dipole 
 responses~\cite{PhysRevB.96.035403}. In principle,
 cross coupling can occur between
 the effective resonators whereby an in-phase
 pair and an out-of-phase pair get
 mixed due to interactions. However,
 here we consider only two
 interacting in-phase pairs, and separately,
 two out-of-phase pairs; neglecting such processes.
 There are two modes of collective
 oscillation for each effective resonator system;
 symmetric and antisymmetric.
 We also designate these as E1s
 and E1a, respectively, when
 the nanorods within each pair are in-phase,
 and E2s and E2a, respectively, when the nanorods within
 each pair are out-of-phase.

 In \Fref{fig:2}, we calculated
 the collective mode decay rates
 for two parallel nanorods. These
 decay rates, $\gamma_{\text{s}}$ and
 $\gamma_{\text{a}}$, provide us the total 
 decay rate for in-phase and out-of-phase pairs
 of nanorods, respectively. Also in \Fref{fig:2},
 we calculated the line shifts of
 the collective modes, this allows us to
 determine the resonance frequencies
 for in-phase and out-of-phase
 pairs of nanorods;
 $\Omega_{\text{s}}$,
 and $\Omega_\text{a}$, respectively.
 In general,
 $\Omega_\text{s}$, $\Omega_\text{a} \ne \omega_0$,
 this means that
 in our effective resonator model, the
 diagonal elements of
 $\mathcal{C}$ also contain the imaginary
 component;
 $\text{Im}\begin{bmatrix}\mathcal{C}\end{bmatrix}_{jj}
 = \delta\omega_\text{s,a}$,
 where $\delta\omega_\text{s,a}$ are the line shifts
 of two in-phase and out-of-phase
 parallel nanorods, respectively,
 see \Fref{fig:2}.

 \begin{figure}[h!]
 \centering
 \includegraphics[]
 {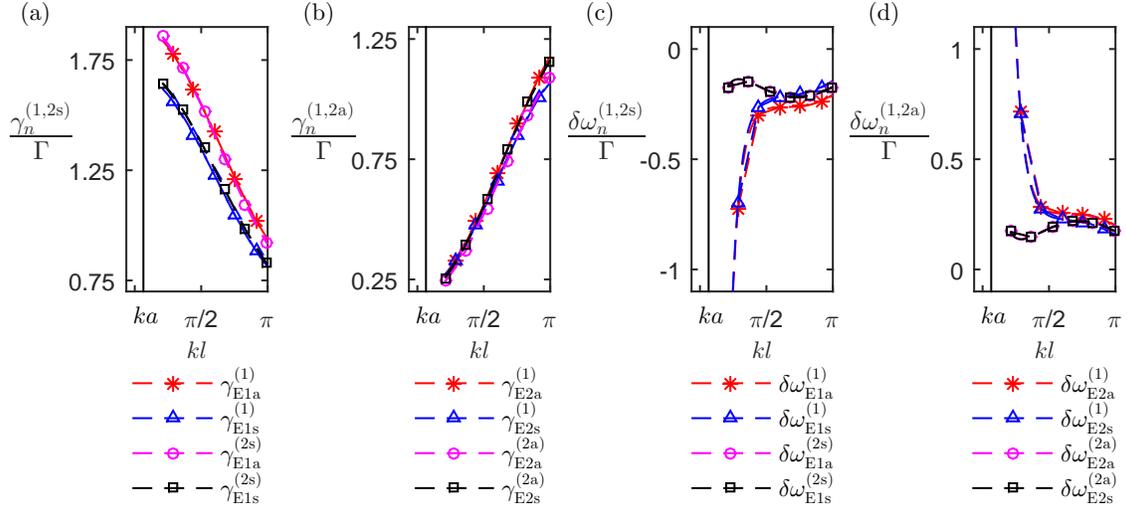}
 \caption
 {\label{fig:5}
 The radiative resonance
 linewidths $\gamma_n^{(\text{1,2s,2a})}$ and
 line shifts $\delta\omega_n^{(\text{1,2s,2a})}$
 for the collective eigenmodes of two horizontal 
 pairs of nanorods and two pairs of point electric dipoles 
 located at ${\bf r}_{\pm,1} = [\lambda_0/2,\pm l/2,0]$ 
 and ${\bf r}_{\pm,2} = [-\lambda_0/2,\pm l/2,0]$ as a
 function of the parameter $l$.
 We show:
 the linewidths $\gamma^{(1)}_{n}$ 
 and line 
 shifts $\delta\omega^{(1)}_{n}$ for the 
 $n=(\text{E1a, E1s, E2a, E2s})$ collective modes
 of the point electric dipole model;
 the linewidths $\gamma^{(2\text{s})}_{n}$ 
 and line shifts $\delta\omega^{(2\text{s})}_{n}$ 
 for the $n=(\text{E1a, E1s})$ collective modes of the
 in-phase effective molecules; and
  the linewidths $\gamma^{(2\text{a})}$ and
 line shifts $\delta\omega^{(2\text{a})}_{n}$, for the 
 $n=(\text{E2a, E2s})$ out-of-phase effective 
 metamolecules.
 For the nanorod parameters
 see \Fref{fig:2} caption.
 }
 \end{figure}

 In \Fref{fig:5},
 we show how the radiative linewidths
 $\big[\gamma_{\text{E1a}}^{\text{(2s)}}$,
 $\gamma_{\text{E1s}}^{\text{(2s)}}$,
 $\gamma_{\text{E2a}}^{\text{(2a)}}$,
 $\gamma_{\text{E2s}}^{\text{(2a)}}\big]$, and line shifts
 $\big[\delta\omega_{\text{E1a}}^{\text{(2s)}}$,
 $\delta\omega_{\text{E1s}}^{\text{(2s)}}$,
 $\delta\omega_{\text{E2a}}^{\text{(2a)}}$,
 $\delta\omega_{\text{E2s}}^{\text{(2a)}}\big]$
 of the collective
 modes of oscillation of the $N=2$ effective interacting pairs
 of nanorods (denoted by the superscripts (2s)  for
 in-phase and (2a) for out-of-phase pairs)
 vary with the parameter $l$, and compare to 
 the corresponding modes' line widths and 
 shifts of $N=4$
 point electric dipole system
 (denoted by the superscript (1)).

 The linewidths
 resulting from the interacting
 $N=4$ point electric dipoles
 closely agree with those of the $N=2$
 effective metamolecule models, both when in-phase
 and when out-of-phase. Also, comparing
 \Fref{fig:5}(a,b), with
 \Fref{fig:4},
 the effective metamolecules' linewidths closely match those
 of the finite-size nanorod model.

 The point electric dipole model's line shifts in
 \Fref{fig:5}(c,d)
 begin to separate from the corresponding
 effective metamolecule line shifts
 when $kl\simeq \pi/2$. As $kl$ reduces, the 
 point electric dipole model's line shifts 
 for the E2a and E2s
 (E1a and E1s) modes
 blue shift (red shift) from the out-of-phase
 (in-phase) effective metamolecule's corresponding 
 modes. Comparing
 \Fref{fig:5}(c,d) with
 the corresponding line shifts in
 \Fref{fig:4},
 the line shifts of the finite-size nanorod model closely
 agree with those of the effective metamolecule model.

 \section*{Conclusions}
 \label{sec:conclusions}

 Understanding the
 complex EM interactions in resonator ensembles is 
 important for the design of metamaterials. Our circuit
 element resonator model provides an efficient way of 
 understanding the dynamics of the system
 without having to fully solve Maxwell's equations.
  We have studied the strong collective modes
 of current oscillations resulting from the 
 EM interactions in closely spaced resonator 
 systems. These  collective modes
 have an associated radiative 
 response that can be either superradiant 
 or subradiant, and together with the
 resonance line shift, is  strongly influenced by the
 spatial separation of the resonators.
 Though in this work, we have considered all the resonance
 frequencies of the resonators to be equal, variation 
 in the resonances (inhomogeneous broadening) 
 can generally suppress the collective radiation 
 interactions~\cite{PhysRevB.86.205128}.

 We have
 analyzed, in detail, the validity of the
 point electric dipole approximation of interacting
 resonators in small systems;
 demonstrating how we can model
 plasmonic nanorod systems both
 as point electric dipole resonators and accounting for their
 finite-size and geometry. In particular,
 we have determined how interacting discrete
 nanorods with an appreciable range 
 of lengths centered on
 $H_0\simeq 0.24\lambda_0 \simeq 210~\text{nm}$
 can be approximated as
 interacting point electric dipoles, especially when
 their separation is greater than $kl\simeq\pi/2$. For 
 closely spaced resonators $kl \lesssim \pi/2$, their finite-size
 and geometry becomes increasingly 
 important.

 An alternative approach for treating each 
 resonator as a separate meta-atom
 is to model closely spaced resonators
 as a single effective metamolecule,
 reducing the number of degrees
 of freedom. In principle,
 this could be extended to other
 more complex effective metamolecules,
 e.g, toroidal metamolecules~\cite{PhysRevB.93.125420}.

 \section*{Acknowledgements}
 We acknowledge discussions with Nikolay I. Zheludev 
 and financial support from the EPSRC and the Leverhulme Trust.

 \section*{Data availability}
 Following a period of embargo, the data from this paper can be
obtained from the University of Southampton ePrints research
 repository: https://doi.org/10.5258/SOTON/D0856.

 \section*{Author contributions statement}

 D.W.W. performed the calculations and prepared the
 manuscript under guidance from J.R. and S.D.J.
 All authors contributed discussion of
 the work and reviewed the manuscript.

 \section*{Additional information}

There are no \textbf{competing interests}.
\end{document}